\documentclass{elsart}
\usepackage{graphicx}
\begin{document}
\begin{frontmatter}
\title{Lattice vibrations in  high-pressure phases
of LiYF$_{4}$}

\author{A. Sen\thanksref{PA}}, 
\author{S. L. Chaplot} and
\author{R. Mittal}
\address{Solid State Physics Division, Bhabha Atomic Research Centre, Mumbai 400085, India}
\thanks[PA]{Present address: Department of Physics, Indian Institute of Technology, Kanpur 208016, India; Electronic address: asen@iitk.ac.in}

\begin{abstract}
Possible variations in the dynamical behaviour of LiYF$_{4}$  due to its several structural changes under pressure are examined by making use of the complementary techniques of quasi-harmonic lattice dynamics and molecular dynamics simulations.  The phonon spectra in the entire Brillouin zone together with the respective Gibbs free energies are calculated for the three high-pressure polymorphs of LiYF$_{4}$ (that are stable at T = 0) with a view to better understand their relative stabilities as functions of pressure and temperature. The present work predicts anomalous thermal expansion at low temperatures in phases I and IIa while irreversibilty of phase II $\rightarrow$ phase III transition on subsequent pressure release. Molecular dynamics simulations provide qualitative  impressions  about a temperature-driven second-order transition and also of kinetic effects in the subsequent pressure-driven first-order phase transformation.
\end{abstract}
\end{frontmatter}

\section{Introduction}
Ternary rare-earth halide  LiYF$_{4}$, which has drawn considerable interest\cite{saran,salaun,senprb1,man,senjpcm1,grez,cur,senprb2,sli,eran} as a laser host material, displays several structural phase transitions\cite{man,grez,senprb2} upon compression. While isostructural CaWO$_{4}$ has a scheelite ($I4_{1}/a$) to wolframite ($P2_{1}/c$) kind of high-pressure phase transformation, LiYF$_{4}$ displays an intermediate fergusonite phase (viz. $I2/a$, above room temperature and $P2_{1}/c$, in the low temperature region) associated with a soft phonon mode\cite{senjpcm1}.  Further, the axial ratio in both the compounds evolves in a different way under the effect of pressure\cite{eran}. Recently, the post-fergusonite phase of LiYF$_{4}$ has been predicted as wolframitelike monoclinic one\cite{senprb2,sli}. 

\vspace{0.2cm}
\parindent 0.8cm
With  the  advent  of  high-energy  synchrotron  X-rays  and sophisticated pressure cells, high-pressure lattice dynamics has gained considerable momentum in recent years  as  it enables a direct comparison of experimental  findings obtained either by the inelastic scattering processes or through some other techniques like ultrafast X-ray diffraction\cite{ultra}. In this work, we  have carried  out  extensive  lattice  dynamical calculations  in the  quasiharmonic  approximation (QHA) for various individual high-pressure phases (e. g. I, IIa and III in Fig.~\ref{fig1}) of LiYF$_{4}$ that may help analyze experimental observations. The  same  interatomic  potential as developed earlier\cite{senprb1} for lattice dynamical calculations under ambient conditions is made  use of here again for high-pressure studies.  Gibbs free energies are calculated for various high-pressure phases of LiYF$_{4}$  in order to have a better understanding of their relative stabilities in terms of volume compression and vibrational entropy. We further examine their respective phonon  dispersions  and  phonon density of states. The second-order phase transitions and the kinetic effects in the first-order phase transformations are specifically investigated by molecular dynamics simulations. Finally, a pressure-temperature (P-T) phase diagram associated with LiYF$_{4}$ is plotted to qualitatively gauge the underlying physics of phase transitions at the microscopic level.

\section{Vibrational Properties}
\subsection{Raman and infrared active modes}
From the previous works\cite{salaun,senprb1,senjpcm1}, we know that the ambient scheelite phase    ($I4_{1}/a$, Z=4) of LiYF$_{4}$ has a total of 36  vibrational degrees of freedom per primitive cell. However, the two high-pressure phases (IIa and III in Fig.~\ref{fig1}) possess  72  phonon  modes at   each wavevector  due  to  doubling  of the primitive cell. Since both these high-pressure phases of LiYF$_{4}$ belong to the same  space group($P2_{1}/c$,  Z=4),  the common irreducible representation of the phonons at the zone center is given by
\begin{center}
              $\Gamma$: 18$A_{g}$+18$A_{u}$+18$B_{g}$+18$B_{u}$,
\end{center}
of  which  A$_{g}$  and  B$_{g}$  modes  are Raman active while modes of ungerade symmetry (viz. A$_{u}$ and B$_{u}$) remain infrared active. Although our calculated high-pressure phases have
36 Raman active modes in all, the ambient fergusonite structure has only 18 such frequencies. It may be  further noted  that the degeneracy of the E$_{g}$ and E$_{u}$ modes which were present
in phase I (scheelite) of LiYF$_{4}$ is lifted in  the high-pressure phases.Table~\ref{tab1} yields a comparison of our calculated zone centre normal modes between phases IIa and III at
respectively 13 and 18 GPa. Phase IIa has a soft A$_{g}$ mode at 18 cm$^{-1}$ which hardens considerably to 92 cm$^{-1}$ in phase III. Most other modes harden slightly in going from phase IIa to phase III. The vibrational frequencies for each of the phases are computed by diagonalization of  the  respective dynamical matrices using the software DISPR\cite{chap}, developed at Trombay.

\subsection{Phonon dispersion relations}
To compare the phonon dispersion in various phases, we resort to the common high symmetry direction of $\Lambda$,  which is  labeled as the $\bf \it c$ axis for phase I ($I4_{1}/a$) while it is the $\bf \it b$ axis for the other high-pressure(monoclinic) phases  of  similar space  group  symmetry (i. e. $P2_{1}/c$).  From our group theoretical calculations, we  obtain  for the irreducible representations of all the normal modes in the three phases as follows:
\begin{center}
Phase I $\rightarrow$ $\Lambda$: 8$\Lambda_{1}$ + 8$\Lambda_{2}$ +
10$\Lambda_{3}$ ($\Lambda_{3}$ being doubly degenerate)\\
Phase IIa and Phase III $\rightarrow$ $\Lambda$: 36$\Lambda_{1}$ +
36$\Lambda_{2}$\\
\end{center}
Phonon dispersions of various polymorphs in LiYF$_{4}$ are displayed in Figure~\ref{fig2} along with the available inelastic neutron scattering data.\cite{salaun} The experimental
data\cite{salaun} for the phase I at zero pressure fit well with the lattice dynamical  calculations\cite{senprb1}. Due to doubling of the unit cell in going from phase I to phase IIa, the Brillouin zone in phase  IIa is halved and so there is an apparent folding back of the dispersion branches from zone boundary (of phase I) to zone center (of phase IIa). This leads to  several optic phonons at the zone center in phase IIa at low energies. We  find a very low  energy zone-center  optic  mode  in  phase  IIa, indicating the onset of a possible high-pressure phase transition at  a pressure  close to 13 GPa. This espouses our previous MD observations\cite{senprb2}. The soft-alike phonon branch of phase IIa may be identified as  the longitudinal A$_{g}$ mode [Table~\ref{tab1}] as noted earlier. A crossover between an  acoustic  phonon  mode and a low frequency optic mode is observed in phase IIa and also in phase III. We further notice that two acoustic phonon branches of $\Lambda_{2}$ symmetry in the phase III are too closely dispersed  to  remain  distinguishable.

\subsection{Phonon density of states}
To have an overview of the range and extent of various phonon modes in all the three phases, we integrate over all the phonons with an energy resolution of 0.5 meV at each wave
vector  on  a  $16\times16\times16$ mesh  within the irreducible Brillouin zone. As Figure~\ref{fig3} suggests, the energy distribution extends by nearly 10 \% in switching from phase  I
to phase  IIa  while the change in the extent is negligible for the phase III transition. In order to check whether  it  is  due to  the pressure effect,  we  compare  the respective frequency distribution for phase IIa and phase III at the same pressure of 13 GPa. One may notice that though the distribution patterns slightly differ, the  extent  remain  quite the
same.  If one is further interested in looking for individual contributions  of  the  constituent  atoms  to  the   entire frequency distribution,  partial  density  of states (PDOS) are to be estimated by atomic projections of the one-phonon eigenvectors.

\vspace{0.2cm}
\parindent 0.8cm
The labeling of the crystallographic axes viz. $a$, $b$, and $c$ varies in high-pressure phases as per convention adopted for specific space groups. In view of these differences, we choose right-handed orthogonal  $x$, $y$ and $z$ axes as follows: $z$ along $\bf c$ and $x$ along $\bf a$ in phase I ($I4_{1}/a$); $z$ along $\bf b$ and $x$ along $\bf c$ in phases
IIa and III ($P2_{1}/c$). Our  calculated  PDOS for various atoms in various phases are portrayed in Figure~\ref{fig4}.  $Y$ atoms are  found  to  contribute  largely  up  to  60  meV  while  $Li$
atoms make it (especially with polarization along  $x$ and  $y$) in the higher energy side of the density of states (DOS). However, $F$ atom contributions are  spread  over  the  entire energy
range  as if to replicate the total DOS to a considerable degree. Such variations in atomic contributions are partly due to the mass effect.

\section{Thermodynamic Properties}
\subsection{Heat capacity and Debye Temperature}
From  the DOS calculations derived out of our simulated data for all  the  three  phases  of  LiYF$_{4}$,  we  compare  the respective  heat  capacities  at  constant pressure(C$_{P}$) as a
function of temperature. Figure~\ref{fig5}(a) demonstrates how C$_{P}$ changes  for  different high-pressure  phases.  We observe  that  although  initially  constant pressure heat capacity for phase III runs  lower,  it goes  up  above  the  room temperature and in the process, even surpasses phase IIa. The reason is that as we go on increasing temperature (T), the difference
C$_{P}$(T) - C$_{V}$(T) [=$\alpha^2_{V}(T)BVT$]  (where C$_{V}$(T) is the constant volume heat capacity and $B$ is the isothermal bulk modulus) becomes significantly large, as inset of
Figure~\ref{fig5}(a) suggests, due to the existence of much higher thermal expansion $\alpha_{V}(T)$ in phase III than that in phase IIa. A detailed discussion on $\alpha_{V}(T)$ follows in the next subsection.

\vspace{0.2cm}
\parindent 0.8cm
We compare the Debye temperature $\theta_D$(T) for the various phases [Figure~\ref{fig5}(b)] as derived from the calculated C$_{V}$(T). It is observed that in the very  low
temperature  region (say,  $<$ 20  K),  $\theta_D$ differs  a  lot from phase to phase (about 14 \% increase for I$\rightarrow$IIa and about 26 \% increase for IIa$\rightarrow$III) while at higher
temperatures (say, 300 K),  it  comes closer  for  phases  IIa and III as difference gets minimal (about 2 \% only).

\subsection{Gr\"uneisen parameter and thermal expansion}
Lattice  anharmonicity,  which  leads  to  a volume dependence of phonon frequencies($\omega_{i}$), is described by the mode Gr\"uneisen parameter\cite{grun}  given as
\begin {eqnarray}
\label{eq1}
\gamma_{i} = - {\partial ln~\omega_{i}\over\partial
ln~V}
\end{eqnarray}
It  is the only kind of anharmonicity that can be taken care of within the framework of quasiharmonic  approximation. Moreover, Gr\"uneisen parameter  is  an important quantity as  it
describes  the thermoelastic behaviour of materials at high pressures and temperatures. It  has  both  the  microscopic  and macroscopic definition, the former relating to the vibrational
frequencies  of  atoms  in  a  material (Eq.~\ref{eq1})  while the latter, to familiar thermodynamic properties such  as  heat capacity  and  thermal  expansion. Unfortunately,   the
experimental  determination of $\gamma_{i}$, defined in either way, is often not so easy, since the microscopic definition requires a detailed knowledge  of the  phonon  dispersion
spectrum of a material, whereas the macroscopic definition  calls for  experimental   measurements   of   thermodynamic properties at  high  pressures  and  temperatures. In this perspective, theoretical model calculation may be of great relief. Figure~\ref{fig6}  displays  our calculated  average $\gamma_{i}$ for various phonon energies and in various phases. It can be seen that below 10 meV, there is a significant number of  modes  with
negative Gr\"uneisen parameter in phase I as well as phase IIa, while phase III modes have all positive $\gamma_{i}$.

\vspace{0.2cm}
\parindent 0.8cm
The effect of pressure on the volume coefficient of thermal expansion($\alpha_{V}$) can be studied through mode Gr\"uneisen parameters($\gamma_{i}$) in the entire Brillouin zone.  In the
quasiharmonic approximation, each phonon mode of energy E$_{i}$ contributes to the thermal expansion by way  of ($1 \over BV$)$\gamma_{i}$C$_{Vi}$, where $C_{Vi}$ denotes the specific heat
(at constant volume) contribution of the $i^{th}$ mode and $V$ the lattice volume. This procedure suits well  to  the ambient fluoroscheelite  system  because  explicit anharmonicity (which
arises mainly out of thermal amplitudes) is not very significant\cite{senjpcm1} compared to the implicit  effect  that involves  phonon  frequency  change  with volume (as  obtained
from $\gamma_{i}$).

\vspace{0.2cm}
\parindent 0.8cm
Figure~\ref{fig7}(a) demonstrates the variation of $\alpha_{V}$ as a function of temperature. It is also apparent that the scheelite as well  as  the initial high-pressure phase of LiYF$_{4}$ have
negative thermal expansion in the low temperature limit (below 100 K). It may  be because $\gamma_{i}$ has large negative values for phonons below 10 meV in these two phases. However, the
third phase has no  such  anomaly, since it has all positive gammas for phonons of all energies (Fig.~\ref{fig6}). The  contributions  to  $\alpha_{V}$ from phonons of
different energies (corresponding to various phases of LiYF$_{4}$) are displayed  in Figs.~\ref{fig7}(b), (c) and (d) respectively. We observe that at low temperatures (e.g. 20 K), contribution of modes upto 10 meV are quite significant to $\alpha_{V}$, but as temperature is increased (e. g. 300 K) higher energy modes get more populated and hence, contribute in a large way to the volume
thermal expansion.

\subsection{Mean squared displacements and thermal anisotropy}
In  order  to  gain  some  insight  over how the phonons of various energies are polarized in various phases of LiYF$_{4}$, we plot (see Figure~\ref{fig8}) the partial contributions of these
phonons to the  mean  squared  thermal amplitude  of  the constituent  atoms.  The mean squared displacement of atom $k$ along $\alpha$ direction is given by
\begin{equation}
\label{eq2}
 U_{\alpha\alpha}(k,T) = \bigl\langle u^2_{k\alpha}
\bigr\rangle_{T} = A{\hbar\over m_{k}} \intop\limits_{0}^{\infty}
{\left(n+1/2\right)\over \omega} g_{k\alpha}(\omega)d\omega,
\end{equation}
where~$n$ = $\left[exp \left(\hbar\omega \over KT\right)-1\right]^{-1}$; $g_{k\alpha}(\omega)$  is  the partial density of states associated with an atom $k$ whose mass is m$_{k}$; $A$ is the normalization constant such that $\int g_{k\alpha}(\omega)d\omega$ = 1. As Figure~\ref{fig8} suggests, the modes at very low energies involve equal displacements  of all  the  atoms  that correspond to the acoustic modes. Interestingly, between 2  and  9 meV, Y and F atoms have larger
amplitudes than what relatively light-weight  Li  atoms  possess. It  may  be  noted  that   the   basic structure (scheelite)  of LiYF$_{4}$ comprises a pair of strongly bonded LiF$_{4}$
tetrahedra and loosely bonded YF$_{8}$ polyhedra\cite{eran}. Larger amplitudes  of  F atoms  in  the  first  two  phases  mark the  presence of librations of the  LiF$_{4}$  tetrahedra.

\vspace{0.2cm}
\parindent 0.8cm
Our calculated values of equivalent isotropic thermal parameters for $Li$, $Y$ and $F$ atoms in the ambient (i. e. P=0) scheelite phase  of LiYF$_{4}$ are found to be 17, 9 and 14 $\times$
10$^{-2}{\rm \AA^{2}}$ as against the respective experimental\cite{garcia}  values of 20, 10 and 17 $\times$ 10$^{-2}{\rm \AA^{2}}$. A detailed comparison  of the anisotropic thermal parameters (U$_{\alpha\alpha}$) at different temperatures among  the different phases of LiYF$_{4}$ is further given in Table~\ref{tab2}. It  may be interesting to note that while $Y$ and $F$ atoms have comparable atomic displacements along the three directons in all the high-pressure phases of LiYF$_{4}$, $Li$ atoms, in contrast, show larger anisotropy along the $z$ direction in phase I, along the $y$ direction in phase II and again along the $z$ direction in phase III.

\subsection{Free energy and phase stability}
A  pressure-temperature phase  diagram  generally portrays various equilibrium phases at constant temperature (T) and  pressure (P)  with  the lowest  Gibbs  free energy (G).   From   phonon   calculations,
the  temperature  dependent vibrational free energy at various hydrostatic  pressures  for  various phases  of  LiYF$_{4}$  can be  estimated in the quasiharmonic approximation.
Thermodynamically, we may write\cite{hill}
\begin {equation}
\label{eq3} G_{n}(P,T) = U_{n} - TS_{n} + P V_{n}
\end{equation}
where U$_{n}$, S$_{n}$~and V$_{n}$ refer respectively to the internal energy, the vibrational entropy and the lattice volume of the ${n}^{th}$  phase. The lattice excitations  are  treated in this work within the quasiharmonic approximation  where  the  full  Hamiltonian  at  a  given
volume   is approximated  by  a  harmonic  expansion  about  the equilibrium atomic positions and anharmonic effects are included through  the  implicit volume
dependence of the vibrational frequencies. Hence, the vibrational  Gibbs  free  energy  (G$_{n}^{vib}$)  is  found  to be satisfyingly  accurate.

\vspace{0.2cm}
\parindent 0.8cm
To  include  vibrational  effects in the present phase diagram, we have calculated dynamical matrices separately  for  each  of  the three phases (viz. I, IIa and III) at  pressure  intervals  of 2 GPa on a $4\times4\times4$ mesh throughout the irreducible Brillouin zone comprising 64 wavevectors. In order to rationalize the  behaviour of  this  dynamical simulation,  we calculate\cite{asen} the enthalpy vs. pressure curves for all the three known structures of
LiYF$_{4}$ and notice that the enthalpy changes (due to internal energy and volume) are predominant in pressure-driven transitions over free energy changes (due to vibrational energy and entropy). Perhaps  this  is  why  most  of  the first-principles  phase diagram   calculations of pressure-driven phase transitions\cite{duca}   do   not   include vibrational  entropies,  though
configurational contributions are always taken into account. The other reason for this may  be  that  vibrational entropy differences  between phases are assumed to be quite small as we
come across in this particular case too.

\vspace{0.2cm}
\parindent 0.8cm
The  stability  of a crystalline phase is largely determined by the minimization of the Gibbs free energy\cite{chapprb}. We have compared the phase-wise free energy and the outcome is shown in Figure~\ref{fig9}. We note from Fig.~\ref{fig9}(a) that at 300 K the free  energy plot  of  the  phase I joins smoothly at 6 GPa to that of phase IIa, which is  consistent  with  the  nature  of  second  order phase transition.  Beyond  8  GPa, phase III has a lower free energy indicating the greater stability of this phase and also the onset  of  a first  order  phase  transition  from  phase IIa to phase III. However, the transition pressure as obtained through the MD simulations\cite{senprb2} is higher than 8 GPa  due to hysteresis.  We observe  the greater stability of phase III at high pressures  arises  primarily  due  to  its  lower  volume, while   the vibrational  energy remains lower and the entropy
becomes higher in phase III providing an additional stability. From the plot of  differences in  vibrational  Gibbs  free energy ($\Delta$G) at 300 K as a function of pressure (Fig.~\ref{fig9}(b)), we find that G for phase III is lower (above 8 GPa) than that for phase IIa. It is also observed (Fig.~\ref{fig9}(c)) that $\Delta$G between phase IIa  and  phase III structures decreases with temperature. Another point to note from the plots of relative free energy vs. temperature (at 8 GPa) is that  in the very  low temperature region (say,  below  100  K), free energy difference shows an anomalous behaviour. It may be attributed to the fact that there is a greater density of low frequency modes in phase IIa than in phase  III (due to the soft phonon modes in the former phase). Table~\ref{tab1} supports this view, where several vibrational modes have been compared between phase IIa and phase III.  At low  temperatures, these  low  frequencies are populated, thereby lowering the  free energy.

\vspace{0.2cm}
\parindent 0.8cm
The above comparison of the Gibbs free energies in various phases provides the phase diagram for the first order phase transitions. However, the second order transitions and also kinetic effects such as hysteresis in the first order transitions are better illustrated through molecular dynamics simulations. A qualitative impression of the phase transitions in  LiYF$_{4}$
is given in Fig.~\ref{fig10} with increasing pressure and temperature (see caption of Fig.~\ref{fig10}). It stands our earlier observation\cite{senprb2} that phase II possesses two different space groups (viz. phase IIa: $P2_{1}/c$ ; phase IIb: $I2/a$) existing in two different ranges of temperaure. The low temperature phase IIa involves small displacements of some of the $F$ atoms from their ideal positions of body-centered symmetry in phase IIb.  The transition IIa $\rightarrow$ III occurs at a much higher pressure of 14 GPa than the equilibrium phase boundary at 8 GPa due to hysteresis at the time scale of the simulation and is found irreversible on release of pressure.

\section{Conclusions}
We have demonstrated in this work that a $\it {lattice}$ $\it {dynamics}$-$\it {molecular}$ $\it {dynamics}$ study can eventually lead one to calculate the vibrational free energy and other essential thermodynamic  functions of a rare-earth ternary halide in various high-pressure phases. Effort has been made to shed light on the understanding of some important physical phenomena associated  with  the phase equilibria.  Further, our results  have the potential ingredients to help  analyze  the experimental inelastic scattering data, if  high-pressure (along with low temperature) phonon measurements are to be carried out in future.

\ack{A.  S.  would  like  to  express his deep sense of gratitude to the Council  of  Scientific  and  Industrial  Research  (CSIR, New Delhi),  India,  for rendering necessary financial assistance
throughout the work and acknowledge  as  well the continuous encouragement and care taken by Dr.   M.   Ramanadham and Dr. V. C. Sahni.}

\newpage
\begin{table}
\caption{\label{tab1}Comparison of calculated zone-center phonon frequencies between phase IIa(P= 13 GPa) and phase III(P=18 GPa) of LiYF$_{4}$ at T=0 K.}
\begin{tabular}{cccccccc}
\hline\hline
\multicolumn{8}{c}{Long wavelength optical phonon modes(cm$^{-1}$)}\\
\multicolumn{2}{c}{A$_{g}$}&\multicolumn{2}{c}{B$_{g}$}&\multicolumn{2}
{c}{A$_{u}$}&\multicolumn{2}{c}{B$_{u}$}\\
Phase IIa&Phase III&Phase IIa&Phase III&Phase IIa&Phase III&Phase IIa&
Phase III\\
\hline
18&92&115&127&0&0&0&0\\
95&107&162&151&80&82&0&0\\
155&151&193&165&87&128&115&142\\
166&175&217&200&185&156&152&183\\
178&214&227&250&196&206&199&195\\
215&225&260&277&266&268&220&205\\
286&271&293&314&293&298&277&230\\
303&313&316&316&310&299&304&259\\
324&317&336&398&343&352&310&278\\
334&363&369&404&358&356&315&337\\
371&377&384&417&408&382&345&368\\
413&410&438&431&434&436&370&383\\
429&436&461&447&455&463&388&410\\
435&443&496&476&463&473&437&435\\
473&463&507&478&495&497&477&495\\
506&481&531&509&514&539&494&526\\
546&562&577&614&578&623&553&552\\
\hline\hline
\end{tabular}
\end{table}

\newpage
\begin{table}
\caption{\label{tab2}Calculated anisotropic thermal parameters (in units of 10$^{-4}{\rm \AA^{2}}$) for  the constituent atoms at various temperatures of 20 and 300 K associated with the three
high-pressure phases of LiYF$_{4}$. It may be noted that in the scheelite phase ($I4_{1}/a$, Z=4), all the $F$ atoms are symmetrically equivalent.  See text for the labeling of $x$, $y$, $z$ directions.} 
\begin{tabular}{ccccccccccc}
\hline\hline
&&\multicolumn{3}{c}{Phase I (P=4 GPa)}&\multicolumn{3}{c}{Phase
IIa (P=13 GPa)}&\multicolumn{3}{c}{Phase III (P=18 GPa)}\\
Sp.&Temp.\\
&(K)&U$_{xx}$&U$_{yy}$&U$_{zz}$&U$_{xx}$&U$_{yy}$&U$_{zz}$&U$_{xx}
$&U$_{yy}$&U$_{zz}$\\
\hline
Li&20&63&63&88&54&72&60&58&64&85\\
&300&137&137&224&98&144&109&95&114&189\\
Y&20&17&17&14&13&12&17&11&12&13\\
&300&107&107&71&63&52&94&39&42&51\\
F(1)&20&41&41&40&36&36&36&29&28&42\\
&300&154&154&124&108&103&112&70&130&106\\
F(2)&20&&&&36&38&42&33&30&34\\
&300&&&&112&113&162&88&88&106\\
F(3)&20&&&&36&36&38&31&36&31\\
&300&&&&110&108&121&78&78&106\\
F(4)&20&&&&35&36&35&34&45&36\\
&300&&&&103&100&113&93&127&106\\
\hline\hline
\end{tabular}
\end{table}

\newpage
\begin{figure}
\caption{\label{fig1}Crystal structures of LiYF$_{4}$ belonging to (a) phase I, (b) phase IIa, (c) phase IIb and (d) phase III.}
\caption{\label{fig2}Comparison  of  phonon dispersion relations
along the same high  symmetry direction  of $\bf\Lambda$  in  the
three  phases  of  LiYF$_{4}$   at respectively  0,  13  and  18
GPa. Lines refer to our calculations while  symbols represent the
experimental data \cite{salaun}.}
\caption{\label{fig3}(a)Calculated phonon density of states, g(E),
in the three phases of LiYF$_{4}$ at respectively 4, 13 and  18
GPa;  (b)  comparison  of g(E) in the two high-pressure phases at
the same pressure of 13 GPa.} \caption{\label{fig4}Calculated
partial  density of states for the constituent atoms with
contributions from polarizations along three orthogonal directions
in  the  three  phases  of LiYF$_{4}$ at respectively 4, 13 and 18
GPa. Since phase I is of tetragonal symmetry,  the  contributions
along  $x$  and  $y$ directions are identical for this phase. The
relationship between the orthogonal directions and the
crystallographic axes is given in section IIC.}
\caption{\label{fig5}Calculated   (a)   heat   capacity (C$_{P}$)
and  (b)  Debye temperature($\theta_{D}$) as a function of
temperature for  the  three  phases  of LiYF$_{4}$  at
respectively  4,  13  and  18  GPa. The inset depicts the
difference of C$_{P}(T)$ and  C$_{V}(T)$ over the same temperature
region.} \caption{\label{fig6}Calculated  average mode Gr\"uneisen
parameter($\gamma_{i}$)as a function of phonon energy for the
three phases of LiYF$_{4}$ at respectively 4, 13 and 18 GPa.}
\caption{\label{fig7}Calculated  volume  thermal expansion
coefficient ($\alpha_{V}$) as a function of temperature for the
three phases of LiYF$_{4}$ at respectively 4, 13 and 18 GPa. (b),
(c) and (d) demonstrate  how  phonons  of  different energies (in
phase I, phase IIa and phase III respectively) contribute to
$\alpha_{V}$ at 20 and 300 K.} \caption{\label{fig8}Calculated
thermal mean squared displacements  (on a semi-logarthmic scale)
of the constituent atoms for (a) phase I, (b) phase IIa and  (c)
phase  III due  to  phonons  of  various  energies at 300 K.
Insets in (a), (b) and (c) refer to the same but on an expanded
scale up to 10 meV for the respective phases of LiYF$_{4}$ at 4,
13 and 18  GPa. It may be noted that in phases IIa and III, there
are four symmetrically different $F$ atoms, however, only the
average contributions of all the $F$ atoms are shown here.}
\caption{\label{fig9}(a)Comparison  of  the  Gibbs  free energy
per atom as a function of pressure in all the three phases of
LiYF$_{4}$, as obtained  from the  lattice  dynamical calculations
at T = 300 K.  The  variation  in  Gibbs  energy difference
($\Delta$G) is plotted in  (b)  with  pressure(P)  and in  (c)
with temperature(T)  giving a qualitative impression of the first
order phase transition that takes place in switching from phase
IIa to a  dynamically more  favourable phase III. Inset in (c)
shows $\Delta$G vs. T up to 200 K on an expanded scale.}
\caption{\label{fig10}Results of phase transitions as observed in
MD simulations runs with increasing pressures at several constant
temperatures (phase I $\rightarrow$ phase II(a/b) $\rightarrow$
phase III) and also with increasing temperatures at several
constant pressures (phase IIa $\rightarrow$ phase IIb). The
transitions among the phase I, IIa and IIb are found to be
reversible while phase III retains on decompression to P=0. The
dashed line indicates the phase boundary between phase II and
phase III as determined by quasiharmonic free energy
calculations.}
\end{figure}

\end{document}